\documentclass[
 groupedaddress,
 amsmath,amssymb, twocolumn, aps, prl, showpacs, showkeys
]{revtex4-1}

\usepackage[latin1]{inputenc}
\usepackage{dcolumn}
\usepackage{bm}
\usepackage[pdftex]{graphicx}
\usepackage{nonfloat}
\usepackage{hyperref}
\usepackage{endnotes}

\newcommand{\Wcm}{\,\mathrm{W/cm}^2}

\newcommand{\mJ}{\,\rm mJ}
\newcommand{\muJ}{\,\rm \mu J}

\newcommand{\eV}{\,\mathrm{eV}}

\global\long\def\e{\mathbf{E}}
\global\long\def\efs{\e_{0}}

\global\long\def\esrf{\e_{\mathrm{crit}}}

\begin{document}

\preprint{APS/surface harmonic generation}

\title{Harmonic Generation from Relativistic Plasma Surfaces in Ultra-Steep Plasma Density Gradients}

\author{C.~R\"{o}del$^{1,2}$}
\author{D.~an der Br\"{u}gge$^3$}
\author{J.~Bierbach$^1$}
\author{M.~Yeung$^4$}
\author{T.~Hahn$^5$}
\author{B.~Dromey$^4$}
\author{S.~Herzer$^1$}
\author{S.~Fuchs$^1$}
\author{A.~Galestian Pour$^1$}
\author{E.~Eckner$^1$}
\author{M.~Behmke$^5$}
\author{M.~Cerchez$^5$}
\author{O.~J\"{a}ckel$^{1,2}$}
\author{D.~Hemmers$^5$}
\author{T.~Toncian$^5$}
\author{M.~C.~Kaluza$^{1,2}$}
\author{A.~Belyanin$^6$}
\author{G.~Pretzler$^5$}
\author{O.~Willi$^5$}
\author{A.~Pukhov$^3$}
\author{M.~Zepf$^4$} 
\author{G.~G.~Paulus$^{1,2}$}

\affiliation{$^1$Institut f\"{u}r Optik und Quantenelektronik, Friedrich-Schiller-Universit\"{a}t Jena, Germany}
\affiliation{$^2$Helmholtz Institut Jena, Germany}
\affiliation{$^3$Institut f\"{u}r Theoretische Physik, Heinrich-Heine Universit\"{a}t D\"{u}sseldorf, Germany}
\affiliation{$^4$Centre for Plasma Physics, School of Mathematics and Physics, Queen's University Belfast, United Kingdom}
\affiliation{$^5$Institut f\"{u}r Laser- und Plasmaphysik, Heinrich-Heine Universit\"{a}t D\"{u}sseldorf, Germany}
\affiliation{$^6$Department of Physics, Texas A\&M University, College Station TX, United States}

\date{\today}

\begin{abstract}
Harmonic generation in the limit of ultra-steep density gradients is studied experimentally. Observations demonstrate that while the efficient generation of high order harmonics from relativistic surfaces requires steep plasma density scale-lengths ($L_p/\lambda <1$)  the absolute efficiency of the harmonics declines for the steepest plasma density scale-length $L_p \to 0$, thus demonstrating that near-steplike density gradients can be achieved for interactions using high-contrast high-intensity laser pulses. Absolute photon yields are obtained using a calibrated detection system. The efficiency of harmonics reflected from the laser driven plasma surface via the Relativistic Oscillating Mirror (ROM) was estimated to be in the range of $10^{-4} - 10^{-6}$ of the laser pulse energy for photon energies ranging from $20-40 \eV$, with the best results being obtained for an intermediate density scale-length.

\end{abstract}


\maketitle
Ultrashort XUV pulses are a promising tool for a wide range of applications including attosecond laser physics and seeding of free-electron X-ray lasers. Typically, they are created by the nonlinear frequency up-conversion of an intense femtosecond driving laser field in a gaseous medium. Remarkable progress has been made to the present date with efficiencies reaching the level of $10^{-4}$ at 20\,nm wavelengths \cite{Kim2008,Sansone2011}. Such efficiencies are not yet available at shorter wavelengths or for attosecond pulse generation and the low intensities at which harmonic conversion takes place in gaseous media, makes harnessing the high peak power in the $0.1 - 1 \rm{PW}$ regime challenging. High-harmonic generation at a sharp plasma-vacuum interface via the Relativistically Oscillating Mirror (ROM) mechanism \cite{Gibbon1996} is predicted to overcome these limitations and result in attosecond pulses of extreme peak power \cite{Tsakiris2006, Gordienko2004}.

While other mechanisms such as Coherent Wake Emission (CWE) can also emit XUV harmonics \cite{Quere2006}, the ROM mechanism is generally reported to dominate in the limit of highly relativistic intensities, where the normalized vector potential $a_0^2=I\lambda^2/(1.37 \cdot 10^{18} \rm \mu m^2 \Wcm) \gg 1$. The efficiency of ROM harmonics is predicted to converge to a power law for ultra-relativistic intensities \cite{Baeva2006}, such that the conversion efficiency is given by $\eta \approx (\omega/\omega_0)^{-8/3}$ up to a threshold frequency $\omega_t\sim\gamma^3$, beyond which the spectrum decays exponentially. Here, $\gamma$ is the maximum value of the Lorentz-factor associated with the reflection point of the ROM process. While these predictions correspond well with the observations made in experiments using pulse durations of the order of picoseconds in terms of highest photon energy up to keV \cite{Dromey2007,Norreys1996} and the slope of the harmonic efficiency \cite{Dromey2006}, no absolute efficiency measurements have been reported to date. 

The plasma density scale-length plays a critical role in determining the response of the plasma to the incident laser radiation. In the picosecond regime, the balance between the laser pressure and the plasma results in the formation of scale-lengths and density profiles which are close to ideal for ROM harmonic generation in terms of efficiency for a broad range of laser pulse contrast. Achieving ultra-short (attosecond) XUV pulses requires lasers with 10s of femtosecond (few-cycle) duration. Under these conditions, there is insufficient time to modify the density scale-length significantly and hence the density gradient and profile become critical control parameters.

Here, we report on the first absolute measurements of the ROM harmonic yield. The highest yield is observed for intermediate pulse contrast, while the yield declines again for the highest pulse contrast, consistent with a plasma vacuum interface approaching step-like conditions. Achieving and verifying such extreme interaction conditions for relativistic laser intensities is an essential step towards exploiting the potential of a wide range of phenomena, such as bright XUV harmonics, radiation pressure driven ion-sources and the formation of relativistic electron sheets \cite{Kulagin2007}.

\begin{figure*}[!ht]
    \includegraphics[width=0.9\textwidth]{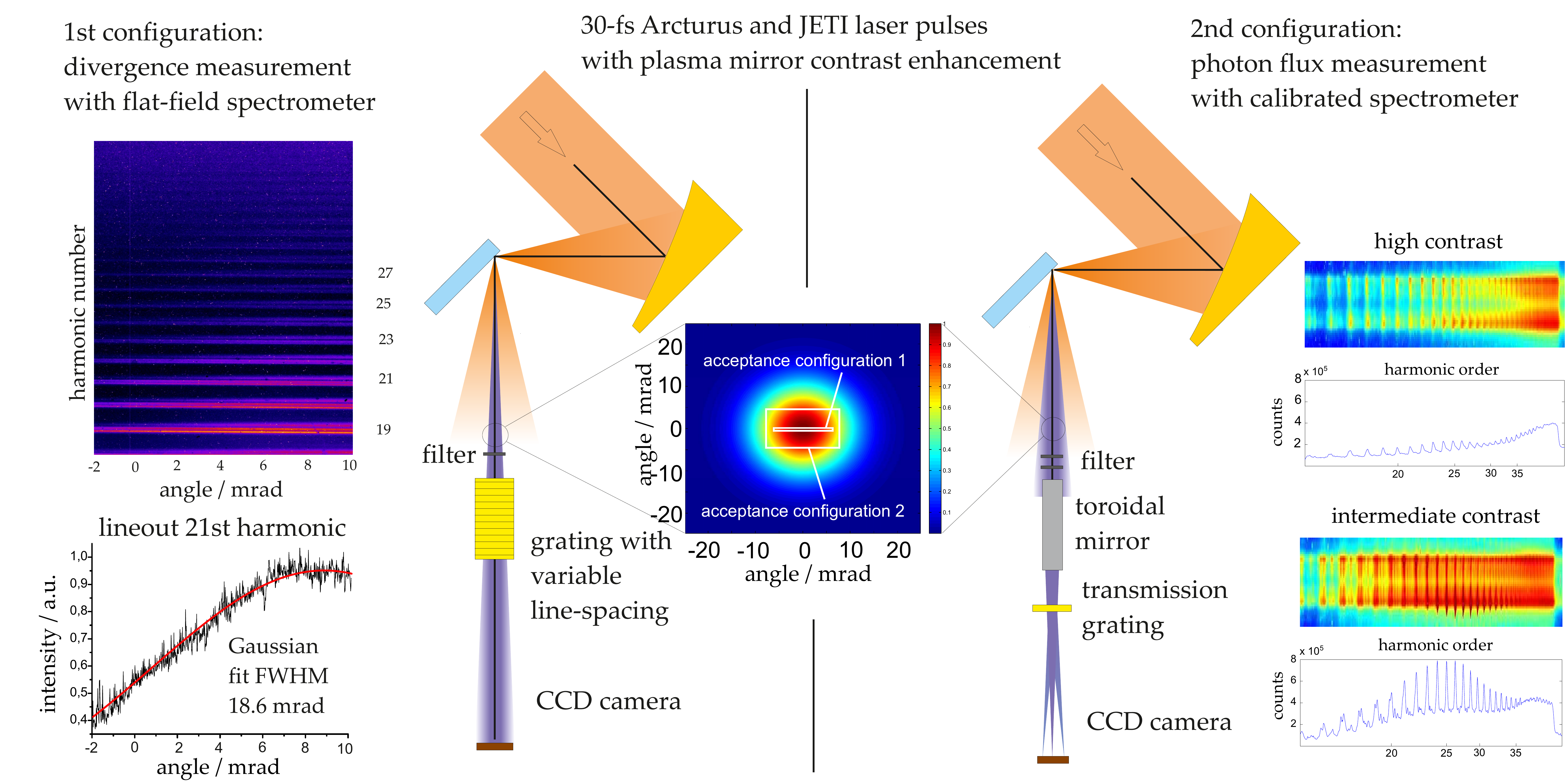}
    \figcaption{Experimental setup: High-contrast laser pulses were focused with an f/2 off-axis parabolic mirror to up to $3 \cdot 10^{19} \Wcm$ on a fused silica or plastic coated substrate at $45^\circ$ p-pol. The XUV emission was recorded with two spectrometers separately (in the presented data only plastic coated targets are used). The flat-field spectrometer shown in configuration 1 allows a measurement of the beam divergence. The XUV spectrometer system in configuration 2 has a larger collection angle and was calibrated regarding the incident photon flux. The black line represents the centroid beam of the laser steering into the center of the XUV spectrometers.}
    \label{setup}
\end{figure*}

Two experiments were performed at the 30-fs Titanium-Sapphire laser systems ``Jeti'' at the University of Jena and ``Arcturus'' at the University of D\"{u}sseldorf, which produced similar harmonic spectra. The laser was focused onto targets made of either glass or photoresist at an incidence angle of $45^\circ$ with an FWHM intensity of $a_0=3.5$. At both laser systems the pulse contrast was controlled by a single plasma mirror with different plasma mirror targets (PMT). The plasma scale length $L_p$ was calculated using the hydrodynamic simulation code ``Multi-fs'' \cite{RAMIS1988} based on the actual pulse profile measured with a 3rd order autocorrelator. Details are given in Ref.\cite{Rodel2011}. The highest pulse contrast was achieved with an anti-reflection (AR) coated PM and resulted in a scale length of $L_p \lesssim \lambda/10 $, while the uncoated borosilicate glass PMs produced an intermediate pulse contrast and $L_p\approx \lambda/5$  \cite{Behmke2011}. The harmonics' pulse energy was determined using an imaging XUV spectrometer with a $8 \,\rm mrad \times 6 \, mrad$ acceptance angle which was calibrated at a synchrotron source \cite{Fuchs2012}. The divergence of the harmonic beam was determined with an angularly resolving XUV spectrometer (see Fig.~\ref{setup} left), which allowed the measured yield to be corrected for the observed angular divergence and thus to obtain the total photon yield. The measurement uncertainty of the photon yield is in the order of 70\,\% (20\,\% spectrometer calibration, 50\,\% uncertainty due to divergence changes). The absolute efficiency $\eta=E_{\rm XUV}/E_0$ was determined by only considering the fraction of the laser pulse energy $E_0$ that contributes to the harmonic generation process. Given the strong non-linearity \cite{Thaury2010} only the fraction of the laser energy that is focused to sufficiently high intensities ($a_0 \geq 1$) contributes to the interaction. Measurements of the intensity distribution of the focal spot with a microscope objective showed $20$ to $40\,\%$ of the laser pulse energy to be concentrated within the FWHM of the focus and thus only this fraction is considered when comparing the measured efficiencies to simulations. 

Two typical harmonic spectra from photoresist targets are shown in Fig.\ref{setup} (right) using high and intermediate contrast settings, respectively. The angular distribution indicated in Fig.\ref{setup} (left) reveals a divergence of $18.6\,\rm mrad$ for the 21st harmonic using photoresist targets and high contrast. The importance of accounting for the divergence of the harmonic beam when comparing changes in the other parameters is highlighted by the observed changes in divergence under conditions where only the pulse contrast was varied. Changing the pulse contrast from the AR to the glass setting (and hence the scale-length from $L_p \approx \lambda/10 $ to $L_p \approx \lambda/5 $) changes the divergence from $18.6\,\rm mrad$ to $26\,\rm mrad$ for the 21st harmonic. In addition, the divergence of different ROM harmonic orders taken from a single measurement has an almost constant value. This is in excellent agreement with previous observations \cite{Dromey2009} and fits well with the analysis that the divergence of the ROM harmonic beam is characteristic of a beam with excellent spatial coherence and is determined by the curvature ('dent') of the emission surface imprinted to the target by the light pressure. Since the velocity of the hole-boring process depends on the plasma density \cite{Wilks1992} the longer scale-length deforms more rapidly and should therefore result in a larger divergence in agreement with the observations. This implies that the divergence should be reduced substantially in the limit of larger spots/or shorter pulses, which would reduce the curvature of the dent at the peak of the pulse respectively.

Until now, it has generally been accepted that achieving sufficiently steep density gradients for ROM harmonics is the major challenge and hence one would expect the harmonic yield to increase as the prepulse level is reduced. For our conditions and peak intensities, the strongest harmonic emission is observed for intermediate contrast settings (glass PMTs), suggesting that the higher contrast setting with AR PMTs has density scale lengths which are even {\it shorter} than the ideal value. The pulse energies $E_{\rm XUV}$ (efficiencies $\eta$) for individual harmonics are in the order of $3-24 \muJ$ ($0.1 -1 \times 10^{-4}$) for the 17th harmonic and approximately $0.3 - 2.7 \muJ$ ($0.1 -1 \times 10^{-5}$) for the 21st using different contrast settings. Thus, we find for the first time that we have clear quantitative evidence of ROM generation in the limit of ultra-steep scale-lengths. While the benefit of a small, but finite, plasma scale-length for ROM has previously been highlighted by simulations \cite{Tarasevitch2007, Thaury2010}, the experiments performed so-far have required the highest achievable pulse contrast or shortest possible scale length, respectively, in order to optimize ROM efficiency and beam quality \cite{Dromey2006,Dromey2009}.

The influence of the plasma scale length has been studied both for glass substrates and photoresist targets that have been coated onto the optically polished glass substrate reducing the density from $2.2 \,\rm g/cm^3$ to $\approx 1.1 \,\rm g/cm^3$ (or from $\approx 400 n_c$ to $\approx 200 n_c$ in terms of the critical density $n_c$). The harmonic emission for these high target densities and respective scale lengths is comparable indicating that the enhanced harmonic emission at intermediate scale lengths is not very sensitive for such high peak densities. This means that for our parameters the harmonic emission is enhanced due to the lower density in the plasma gradient and \textit{not} by using a lower maximum density. Since the reflection point of the ROM is located near the critical density at elongated plasma density ramps, the ROM process is affected by the length of the plasma gradient instead of the maximum plasma density. The observed dependence of the efficiency on the scale length can be understood in terms of the plasma dynamics as follows.  First, the denser the plasma and the steeper the gradient, the more the electric field in the skin layer is reduced. Second, the ``spring constant'' of the electron plasma becomes larger for denser and steeper plasmas, making the ROM harder to drive to the high values of $\gamma$ associated with a more efficient production of higher harmonic orders.

\begin{figure}[h]
    \includegraphics[width=3.375in]{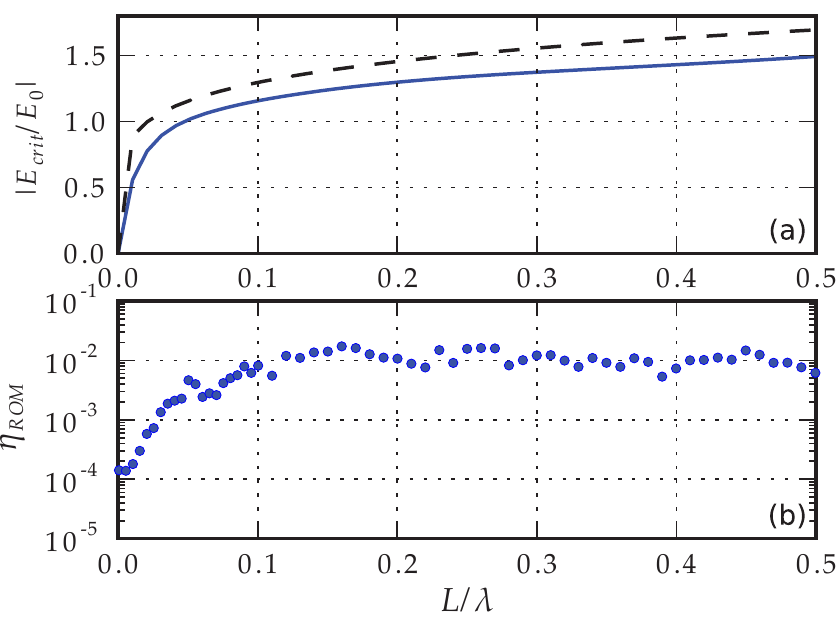}
    \figcaption{(a) Surface field $\esrf$ at the critical density in units of the incident field $\efs$,
    as estimated from the equation in Ref.~\cite{Kruer2003} (black dashed line) and computed exactly by numerical
    integration for an exponential gradient (blue line).
    (b) Efficiency of SHHG above the 14th order
    $\eta_{ROM} = \int^{\infty}_{14 \omega_0} I(\omega)d\omega / P_0$
    for $a_0=3.5$ at different plasma scale lengths from a set of 1D PIC simulations.
    Incidence was p-polarized, the plasma ramp is exponential up to a maximum density of $n_e = 200 n_c$. }
    \label{scalings}
\end{figure}

To make an analytical estimate of the field at the critical density surface we can consider a laser interacting at normal incidence with the target (oblique incidence can be treated by switching into the frame of reference, in which the laser is normally incident \cite{Bourdier1983}). The electrons can gain kinetic energy only through the $\e$-field of the laser. This field is tangential to the surface and is attenuated due to the skin effect. We assume for the moment, that the field is non-relativistic. Evaluating the linear wave equation, we find the threshold condition for it to become relativistic. For a perfectly steep plasma edge, it can be calculated analytically by evaluation of the continuity condition at the plasma edge, yielding $|\esrf|/|\efs|=2\omega_{0}/\omega_{p}$. Hence, for our laser and plasma parameters, the field would not be relativistic for a perfect step density profile. This is reflected in Fig.~\ref{scalings}(b): In the limit $L\rightarrow 0$ there are no relativistic harmonics. The skin field is however enhanced due to a finite density ramp. As a first estimate for finite ramps we may consider the calculation found in Ref.~\cite{Kruer2003}, leading to $|\esrf|/|\efs| \approx 1.4 (\omega_0 L / c)^{1/6}$ at the critical density. This formula becomes exact for linear and extended ($L\gg\lambda$) gradients. For steep, exponential ramps, as are expected in the experiments, we find the skin field by numerical integration of the inhomogeneous wave equation. Results of this computation are shown in Fig.~\ref{scalings}(a), along with the simple scaling from Ref.~\cite{Kruer2003}. It can be seen that even small scale lengths can considerably boost the skin field compared to the case of step-like profiles. Already at $L=\lambda/20$, there is practically no attenuation at the critical density, but the field can still grow slightly for longer plasma scales. We further note that the simple sixth-root dependence calculated for a linear gradient also yields a reasonable estimate for the exponential gradient, only slightly overestimating the field in comparison to the exact result. Fig.~\ref{scalings}(b) shows the integrated efficiency $\eta_{\mathrm{ROM}}$ of ROM harmonics for the same density gradients and a laser amplitude $a_0=3.5$. As expected from the previous considerations and Fig.\ref{scalings}(a), it can be seen that the ROM efficiency rises quickly as soon as the skin field becomes relativistic. For the scale lengths $L>\lambda/10$, the integrated efficiency remains approximately constant at $\eta \approx 7\times10^{-3}$ as expected from the ROM model. While the reduction in the critical field explains the drop to very low efficiencies at very short scale-lengths, it does not fully explain the efficiency scaling at intermediate scale-lengths.

Since the reflection point oscillates around the immobile ion background due to the driving relativistic laser field, another contribution must come from the restoring force due to the quasi-static field generated by the plasma once the electrons are driven out of equilibrium by the laser field. For a given mean displacement of the plasma electrons, the restoring force is proportional to the plasma density. In the limit of a step-like density profile, the restoring force is directly proportional to the maximum plasma density, while in the limit of very long density scale-lengths the restoring force is determined by the critical density. In the intermediate case of relevance here, the restoring force will depend in a complex fashion on the density scale-length, peak density and amplitude of the oscillation. While it is not easily possible to express this dependence in a closed analytical form, it is clear that one would expect the effective density and hence the restoring force to be lower for increasingly shallow density gradients resulting in the dependence shown in Fig.~\ref{scalings}. Generally, the denser the plasma and the steeper the density ramp, the harder it is for the laser to drive large amplitude oscillations at the plasma surface. This in turn leads to a smaller oscillation amplitude of the ROM and, consequently, to a lower $\gamma$-factor. An analysis of the electron spring model at ultra-relativistic intensities can be found in Ref.\cite{Gonoskov2011}. In agreement with our experimental observations a trend towards higher efficiency for moderately long scale-length or low peak densities is expected.

\begin{figure}[!ht]
    \includegraphics[width=0.5\textwidth]{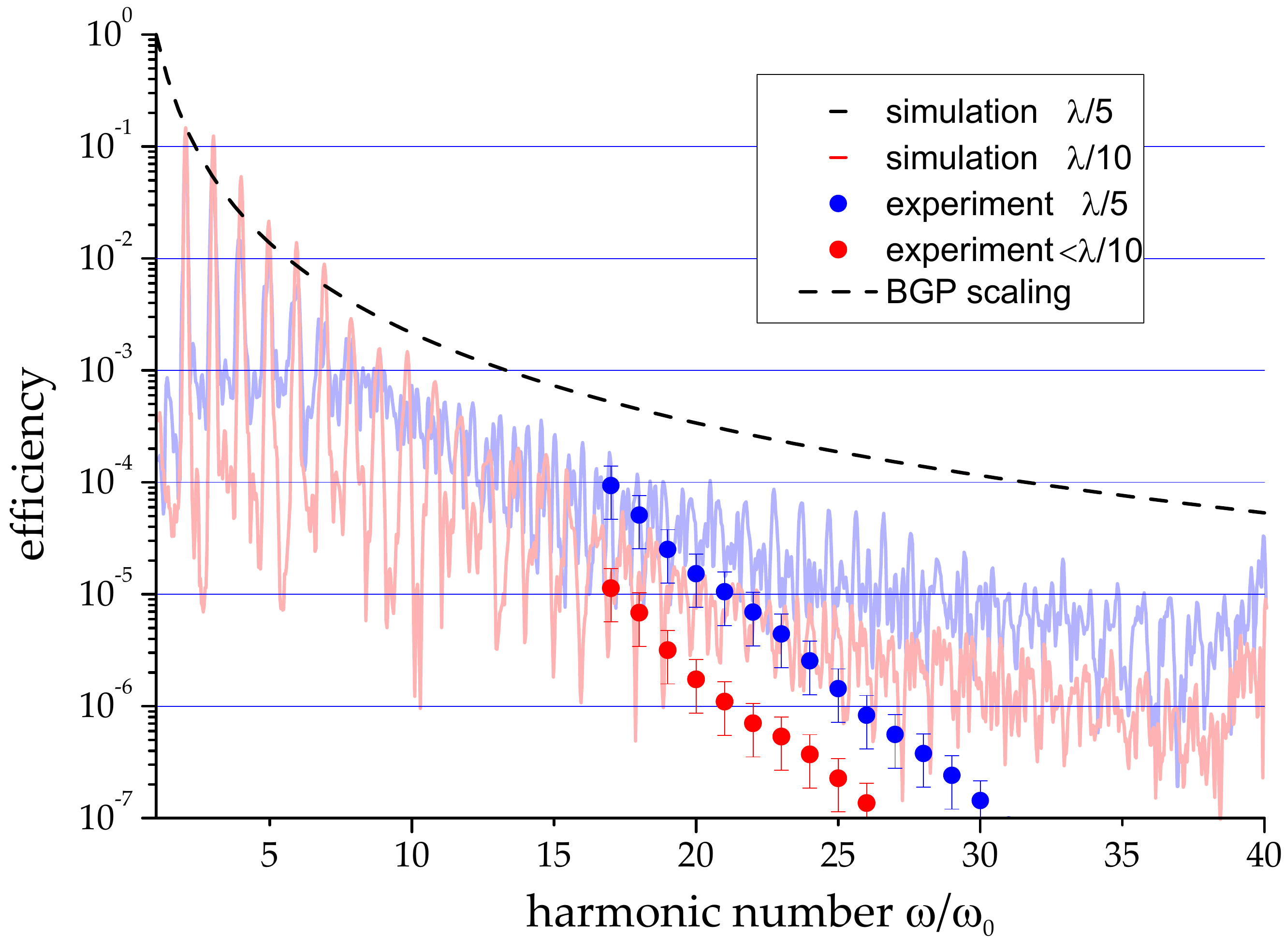}
    \figcaption{Experimental efficiencies (circles) are compared to spectral densities from 1D PIC simulations (lines) for different plasma scale lengths (density $n_e=200n_c$, exponential density profile). The experimental efficiencies have been normalized to a pulse energy of $250 \mJ$ (energy that is focused to $a_0>1$). In the ultra-relativistic limit the efficiencies converge to the BGP power scaling $\eta \approx (\omega/\omega_0)^{-8/3}$ \cite{Baeva2006}.}
    \label{efficiency}
\end{figure}

Under our experimental conditions ($a_0=3.5$) the efficiencies are still expected to be below the relativistic limit regime where the $\eta \approx (\omega/\omega_0)^{-8/3}$ scaling applies. Fig. \ref{efficiency} compares our experimental results to a range of efficiencies predicted by 1D PIC simulations. While the efficiencies are broadly compatible with the range of efficiencies predicted by the simulations, they appear somewhat lower than predictions for the nominal density gradients derived from the measurement of the pulse-contrast and Multi-fs modelling. While, to our knowledge, Multi-fs is the best suited code to calculate the hydrodynamic expansion under such conditions, the code has not been validated directly by measurements of the scale-length under such conditions. Consequently one possible explanation for the discrepancy may be that the density gradients are even steeper than predicted. What is clear both experimentally and from simulations is that the efficiency of the ROM process depends sensitively on the plasma scale length. The generation of surface waves, which have been found in 2D simulations, induce high harmonic emission at angular sidebands \cite{Brugge2012}. This may lead to differences between the experimental results and the 1D simulations. Another important effect that is not considered in our simulations is the ion motion. In fact, Thaury and Quéré \cite{Thaury2010} have shown that the harmonics efficiency in simulations with mobile ions is significantly reduced.

In conclusion, we have investigated harmonic generation in the limit of ultra-steep density gradients and shown first experimental evidence of the absolute yield reducing for very steep gradients. This demonstrates that relativistic interactions in the limit of ultra-steep density gradients can be achieved by a careful control of the laser parameters. Harmonic efficiency is optimized for intermediate scale-lengths. Our results suggest the generation of intense attosecond pulse trains with pulse energies exceeding $10 \muJ$, thus paving the way towards applications such as nonlinear attosecond experiments or the seeding of free-electron lasers with surface high-harmonic radiation.

\begin{acknowledgments}
This work was funded by the DFG project SFB TR18 and Laserlab Europe. C.R. acknowledges support from the Carl Zeiss Stiftung. Monika Toncian, Burgard Beleites and Falk Ronneberger contributed to this work by operating the Arcturus and Jeti laser facility.
\end{acknowledgments}

\bibliographystyle{unsrt}
\bibliography{papers}

\end{document}